\documentstyle [12pt,twoside]{article}
\newskip\humongous \humongous=0pt plus 1000pt minus 1000pt
\def\caja{\mathsurround=0pt}

\newif\ifdtup
\def\panorama{\global\dtuptrue \openup2\jot \caja
        \everycr{\noalign{\ifdtup \global\dtupfalse
        \vskip-\lineskiplimit \vskip\normallineskiplimit
        \else \penalty\interdisplaylinepenalty \fi}}}

\def\eqalignnotwo#1{\panorama \tabskip=\humongous
        \halign to\displaywidth{\hfil$\displaystyle{##}$
        \tabskip=0pt&$\displaystyle{{}##}$
        \tabskip=\humongous&\llap{$##$}\tabskip=0pt
        \tabskip=0pt&$\displaystyle{{}##}$\hfil
        \crcr#1\crcr}}

\def\begintitle#1#2#3#4
        {\begin{titlepage}
         \centerline{#1 \hfill UMDGR-92-#2}
         \begin{center}\vglue .4in
         {\large\bf #3}\\[.4in]
         {\bf #4}\\[.2in]
         {\it Department of Physics}\\
         {\it University of Maryland, College Park, MD 20742}\\[.4in]
         {\bf ABSTRACT}\\
         \end{center}
         \begin{quotation}}
\def\endtitle
         {\end{quotation}
          \end{titlepage}
          \newpage}

\def\cA{{\cal A}}

\def\cP{{\cal P}}

\def\a{\alpha}\def\g{\gamma}\def\e{\epsilon}

\def\u{\mu}\def\v{\nu}\def\s{\sigma}\def\t{\tau}
\def\w{\omega}
\def\L{\Lambda}\def\S{\Sigma}

\def\ovr{\overline}

\def\utw#1{\rlap{\lower1ex\hbox{$\sim$}}#1{}}
\def\pb#1{\rlap{\lower1ex\hbox{$\leftarrow$}}#1{}}
\def\pf#1{\rlap{\lower1ex\hbox{$\rightarrow$}}#1{}}
\def\X{\times}

\def\3#1{{}^3\!#1}\def\4#1{{}^4\!#1}\def\+#1{{}^+\!#1}\def\-#1{{}^-\!#1}
\def\*#1{{}^*\!#1}

\begin{document}

\begintitle{May 1992}{208}{\hfil THE SPIN HOLONOMY GROUP\hfil\break
\centerline{IN}\hfil\break GENERAL RELATIVITY}{Ted
Jacobson\footnote{jacobson@umdhep.umd.edu} and Joseph
D. Romano\footnote{romano@umdhep.umd.edu} }
It has recently been shown by Goldberg {\it et al} that the holonomy
group of the chiral spin-connection is preserved under time evolution
in vacuum general relativity. Here, the underlying reason for
the time-independence of the holonomy group is traced to the self-duality
of the curvature 2-form for an Einstein space.
This observation reveals that the holonomy group is time-independent
not only in vacuum, but also in the presence of a cosmological
constant. It also shows that once matter is coupled to gravity,
the ``conservation of holonomy" is lost.

When the fundamental group of space is non-trivial, the holonomy group need not
be connected. For each homotopy class of loops, the holonomies comprise a
coset of the full holonomy group modulo its connected component. These cosets
are also time-independent. All possible holonomy groups that can arise
are classified, and examples are given of connections with these holonomy
groups. The classification of local and global solutions with given holonomy
groups is discussed.

\vspace{.4cm}


\endtitle

\noindent{\bf 1. Introduction}
\vskip .5cm

Since there is such a dearth of
known observables in general relativity, any observable is worth studying.
This is especially true in view of issues raised by quantum
theory. For example, it is only when true observables are known that
the physical inner product in Hilbert space can be constrained by
reality conditions, and meaningful physical statements can be extracted
from the theory. Moreover, an observable constructed entirely from the
chiral spin-connection is particularly interesting because, as realized
by Ashtekar, the components of this connection form a complete set of
coordinates having vanishing Poisson brackets on the phase space of
complexified general relativity.  The corresponding quantum operators
therefore commute, so an observable built purely from them may be free
of operator-ordering ambiguities.

It is therefore noteworthy that the holonomy group of the chiral
spin-connection is an observable in vacuum general relativity (GR).
More precisely, as shown in a recent paper by Goldberg, Lewandowski,
and Stornaiolo \cite{GLS}, the complexification of the holonomy group based at
a point $*$ is preserved under time evolution. In addition, this holonomy
group is invariant under spatial diffeomorphisms and $SL(2,C)$
spin-transformations that are the identity at $*$. Thus, the holonomy group
qualifies as an observable once the basepoint $*$ and spin-frame at
$*$ are fixed. This observable is determined by the spin-connection on an
initial value hypersurface.  Thus, in a hamiltonian formulation of the theory,
it is determined by the phase space coordinates, without implicitly or
explicitly solving the dynamics.

In this paper, the result of \cite{GLS} is extended in several directions.
First, the underlying reason for the time-independence of the holonomy group
is traced to the self-duality of the curvature 2-form in the absence of matter.
This observation reveals that the holonomy group is time-independent
not only in vacuum, but also in the presence of a cosmological constant. It
also shows that once matter is coupled to gravity,  the ``conservation of
holonomy" is lost.

For a generic point in phase space, the holonomy group will
be all of $SL(2,C)$. Thus, the holonomy group observable
does not contain very much information about the gravitational field in
general, and for this reason, it would appear not to be a very interesting
observable. It is, however, better than nothing. Moreover, when the
fundamental group of space is non-trivial, there is a refinement of the
holonomy group observable. For each homotopy class of loops, the holonomies
comprise a coset of the full holonomy group. This coset is also
time-independent. But since the fundamental group is not, in general,
invariant under large diffeomorphisms,
the homotopy-class holonomy cosets are not quite observables. To obtain
observables, the action of these diffeomorphisms must be factored out.

This construction mirrors that of 2+1 dimensional gravity, where the
vacuum equations imply local flatness of the $SO(2,1)$ frame-connection,
and the map from homotopy classes to holonomy elements yields an observable
after factoring by the mapping class group \cite{witten,observables}.
Remarkably, the 3+1-dimensional homotopy observables exist even though the
connection is not necessarily flat. They are non-trivial, however, only if the
holonomy group is not all of $SL(2,C)$.

The rest of this paper is organized as follows: First, in section 2, we provide
our alternate proof of the conservation of holonomy, showing that the
result extends to the case of a cosmological constant, but not to arbitrary
matter coupling. Next, in section 3, we spell out the relation between this
local result and the global statement that the holonomy group is an observable.
This discussion is intended to provide an explicit treatment of some points
that were implicit in \cite{GLS}. In order to have
the most general result, we will take care to allow for arbitrary spatial
topology and arbitrary connections. In section 4, the homotopy observable is
introduced, and in section 5, we classify the cases that can potentially arise
for this observable. In section 6, the local classification of solutions
with restricted holonomy algebras is given, and the global classification
problem is discussed but not solved. Finally, section 7 contains
a brief discussion of the results, their possible uses, and open questions.
An appendix contains a proof of the reduction theorem used in section 3.

\vskip 1cm
\noindent{\bf 2. Local Result: Time-independence of the connection algebra}
\vskip .5cm

The local result was proved in \cite{GLS} using Ashtekar's hamiltonian
formulation of GR, in a case by case analysis of the subalgebras of the Lie
algebra $sl(2,C)$. It turns out to be much easier to see what is happening by
taking a more 4-dimensional point of view.  In what follows, we will assume
that the spacetime manifold $M$ has the form $M=\S\X R$ for some connected
3-manifold $\S$, and that the spacetime metric is non-degenerate.
Degenerate metrics were considered in \cite{GLS} using the Ashtekar variables
approach.  In general, time independence of the holonomy
group is lost when the metric is degenerate, although there are special cases
for which it still holds.

Consider a 4-dimensional left-handed spin-connection $\w_\u{}^{AB}$ and
its curvature 2-form $R_{\u\v}{}^{AB}$ at any spacetime point $(x,t)$
and in any gauge. If $\w_\u{}^{AB}$ is the spin-connection corresponding to
a vacuum solution of Einstein's equation, then the vanishing of the tracefree
part of the Ricci tensor implies that the curvature 2-form $R_{\u\v}{}^{AB}$
is self-dual.  In coordinates that are orthonormal at a point, this means that
$$R_{0i}{}^{AB}=i\e^{ijk}R_{jk}{}^{AB},\eqno(1)$$
where $i,j,k$ are spatial indices and $0$ is timelike, and we have assumed
Lorentz signature.

In a gauge with $\w_0{}^{AB}=0$, \ $R_{0i}{}^{AB}$ is simply the time
derivative of $\w_i{}^{AB}$. If in a neighborhood of $x\in\S$, $\w_i{}^{AB}$
initially takes values in some subalgebra of $sl(2,C)$, then $R_{ij}{}^{AB}
(x)$ takes values in the same subalgebra.  Thus, under time evolution
according to the vacuum equations, the self-duality equation (1)
shows that for some time interval, $\w_i{}^{AB}(x)$ will remain in this
subalgebra in a gauge with $\w_0=0$.\footnote{After a finite time interval has
passed, it may happen that there is no longer a neighborhood of $x$ in
which $\w_i{}^{AB}$ falls within the subalgebra, since curvature can
propagate to $x$ from other regions.  If this happens, the connection will go
outside the original subalgebra \cite{lewandowskipc}.
Nevertheless, as will be explained
in section 3, the complexification of the full holonomy {\it group}
will be conserved.} Note that because of the factor of $i$ in equation (1),
it is the {\it complex} subalgebra that is preserved. Note also that this
argument actually shows that the holonomy algebra is preserved along
{\it any} foliation of spacetime, not just a spacelike one.

Since the curvature 2-form $R_{\u\v}{}^{AB}$ is also self-dual in the
presence of a cosmological constant, the corresponding subalgebra is
time-independent in a gauge with $\w_0=0$ in that case as well. However, in
the presence of matter, $R_{\u\v}{}^{AB}$ is no longer self-dual. Then
we have
$$R_{\u\v}{}^{AB}=X^{ABCD}\S_{\u\v CD}+\Phi^{ABC'D'}\ovr\S_{\u\v C'D'},\eqno
(2)$$
where $\S_{CD}=\theta_C{}^{C'}\wedge\theta_{DC'}$, with \ $\theta_\u{}^{CC'}$
being tetrad 1-forms. The 2-forms $\S_{CD}$ are self-dual, and the conjugates
$\ovr\S_{C'D'}$ are anti self-dual. These are independent, and together they
span the 6-dimensional space of 2-forms. $\Phi^{ABC'D'}$ is the spinor
equivalent of the tracefree part of the Ricci tensor, which according to
Einstein's equation is proportional to the tracefree part of the
energy-momentum tensor.  Now if $R_{ij}{}^{AB}$ takes values in some subalgebra
on an initial value surface, this will in general no longer imply anything
about
$R_{0i}{}^{AB}$, since $R_{\mu\nu}{}^{AB}$ has independent self-dual and anti
self-dual parts.

\vskip 1cm
\noindent{\bf 3. Global Result: The holonomy group observable}
\vskip .5cm

The local result just proved states that the complex subalgebra spanned by the
values of the connection 1-form $\w_i{}^{AB}(x)$ at a point $x\in\S$ in a gauge
with $\w_0=0$ is temporarily time-independent in vacuum GR.\footnote{See the
previous footnote.} But this result does not directly yield an
observable. First of all, ``temporary" time-independence is not good enough!
Moreover, the above subalgebra is not invariant under spatial diffeomorphisms,
since it depends essentially arbitrarily on the choice of $x$. Both of
these problems can be remedied by considering the smallest subalgebra
containing the union of the subalgebras spanned by $\w_i{}^{AB}(x)$ for all
$x\in\S$, in some globally defined gauge.\footnote{A global gauge always
exists, since $SL(2,C)$ bundles over 3-manifolds are always trivial. To see
why, note that the obstructions to trivializing a $G$-bundle are the homology
groups $H_k(\Sigma,\pi_{k-1}(G))$. The first non-trivial homotopy group of
$SL(2,C)$ is $\pi_3$, but $H_k$ of a 3-manifold $\S$ vanishes for $k>3$.}

The resulting union of subalgebras is still not an observable, however,
because it is not invariant under arbitrary {\it local} $SL(2,C)$
transformations. Only if the resulting subalgebra happens to be an ideal will
it be invariant. Thus, one way to obtain an observable is to take the smallest
ideal generated by the span of $\w_i{}^{AB}(x)$ for all $x\in\S$.  This raises
the question, what ideals of $sl(2,C)$ are there? (We are interested in complex
ideals, since the result of time-independence applies only to the
complex subalgebra.) Unfortunately, the only ideals  of $sl(2,C)$ are the
identity and the whole algebra, so this observable only distinguishes flat
from non-flat connections.

\vskip .5cm
\noindent{\sl The Holonomy group}
\vskip .25cm

A more interesting observable, as noted by Goldberg {\it et al} \cite{GLS},
is given by the holonomy group of the connection, which is defined as
follows: A basepoint $*\in\S$ is fixed, and one considers all closed
loops in $\S$ that begin and end at $*$. The holonomy $h_\g[\w]$
of the connection 1-form $\w$ around a loop $\g$ based at $*$, is the $SL(2,C)$
element determined by the parallel transport of a spin-frame (or gauge) at $*$
around the loop. If the connection is given in a global gauge, then the
holonomy has the standard form, $h_\g[\w]=\cP\exp\oint_\g\w$, where $\cP$
indicates path ordering. The set $\{h_\g[\w]|\; \hbox{\rm $\g$ is a loop based
at $*$}\}$ is a subgroup of $SL(2,C)$, called the holonomy group based at $*$.
Its Lie algebra is called the holonomy algebra based at $*$.

If the loop $\g$ is covered by a finite number of local gauges
$\s_0,\s_1,\cdots,\s_n$, then the holonomy element can be expressed as
$$h_\g[\w]=(\cP\exp\int_{y_{n0}}^{*}\w^{(0)})(\psi_{n0})^{-1}\cdots
(\cP\exp\int_{y_{01}}^{y_{12}}\w^{(1)})(\psi_{01})^{-1}
(\cP\exp\int_{*}^{y_{01}}\w^{(0)}),
\eqno(3)$$
where $y_{ij}$ is any point in the overlap region of the local gauges
$\s_i$ and $\s_j$, $\w^{(i)}$ is the connection in the gauge $\s_i$,
and $\psi_{ij}$ is the gauge transformation
({\it i.e.}, transition function) at $y_{ij}$ from $\s_i$ to $\s_j$.
To see that (3) is the appropriate expression for the holonomy, note that
under a gauge transformation at the endpoints, the parallel propagator
transforms as $\cP\exp\int_a^b\w'=g_b^{-1}\, (\cP\exp\int_a^b\w)\, g_a$.
Thus, if $\s_0$ is a global gauge and if all the propagators in (3)
are expressed in the gauge $\s_0$ using this transformation formula, then
(3) collapses to the standard form, $\cP\exp\oint_\g\w^{(0)}$.

The holonomy group based at $*$ is invariant under gauge
transformations that are the identity at $*$. It is also invariant under
spatial diffeomorphisms that leave $*$ fixed. This is because such
diffeomorphisms merely mix up the various loops based at $*$.\footnote{In the
hamiltonian picture, the diffeomorphisms act on the spin-connection, which is
a phase space variable, and leave the points of the submanifold $\S$ fixed.
But the holonomy of the transformed connection around a loop $\g$ equals the
holonomy of the original connection around the loop that is mapped
into $\g$ via the diffeomorphism.} The holonomy group is conjugated by some
group element if the gauge at $*$ is changed or if a different basepoint
and gauge are chosen \cite{thebook}. This residual gauge and diffeomorphism
dependence of the holonomy group is typically regarded as rather
benign, since it involves only a single overall conjugation.
To eliminate this dependence, it would be necessary to tie both $*$ and the
gauge at $*$ to some other physical quantity.

\vskip .5cm
\noindent{\sl Time-independence}
\vskip .25cm

Our next step is to show that the local result discussed previously implies
that the complexification of the holonomy group based at $*$ is preserved
under time evolution. (Here the complexification of a subgroup $H$ of
$SL(2,C)$ is defined as the smallest subgroup containing $H$ and the
exponential of the complexification of the Lie algebra of $H$.) The most
elegant way to show this is to invoke the {\it reduction theorem} for a
connection on a principal fiber bundle \cite{thebook}. The reduction theorem
states that the set of points that can be joined to a point $p$ in the
principal fiber bundle by a horizontal curve forms a subbundle, with structure
group equal to the holonomy group at $p$. (This subbundle is called
the holonomy bundle through $p$.) The reduction theorem implies that $\S$ can
be covered by a collection of local gauge patches such that ($a$) in each
gauge patch
the connection takes values in the holonomy algebra based at $*$, and
($b$) the gauge tranformations relating the patches take values in the
holonomy group based at $*$. For completeness, we give a simple proof of the
reduction theorem in the appendix.

Now condition ($a$) and the local result together imply that there exists a
set of local gauge patches covering the spacetime $M=\S\X R$, in which the
connection remains within the original holonomy algebra based at $*$. These
gauges are obtained by imposing the additional gauge condition $\w_0=0$ to
extend the local gauges on $\S$ to the spacetime $M$. With this gauge
condition, the transition functions are time independent, so condition ($b$)
ensures that for this set of extended gauges, the transition functions all
lie within the original holonomy group based at $*$. If we use these results
in conjunction with expression (3) for the holonomy of the connection around a
loop $\g$ based at $*$, we see that under time evolution, the holonomy group
never goes outside the complexification of the original holonomy group. In
fact,
nor can the holonomy group shrink to a proper subgroup of the original
group, since the Einstein equation is time reversal invariant, and the
time-reversed process would violate the previous result. In other words,
the complexification of the holonomy group is preserved under time evolution.

\vskip 1cm
\noindent{\bf 4. Homotopy Observables}
\vskip .5cm

If $\S$ is not simply connected, each homotopy class of loops determines a
collection of holonomy elements, and this collection forms a {\it coset} of
the full holonomy group modulo its connected component. This gives rise to a
refinement of the holonomy group observable, as will now be explained.

Let $\Phi$ be the full holonomy group
(at a basepoint $*$ in a fixed gauge at $*$), and let
$\Phi_0$ be the restricted holonomy group, {\it i.e.,} the group of holonomies
around contractible loops. $\Phi_0$ is a connected, normal subgroup of $\Phi$,
and the quotient group $\Phi/\Phi_0$ is countable. It follows that $\Phi_0$
is, in fact, the same as the connected component of the identity in $\Phi$.
The holonomy map from the loop group to $\Phi$ passes to a homomorphism
$$f\!:\!\pi_1(\S)\rightarrow\Phi/\Phi_0,\eqno(4)$$
from the fundamental group $\pi_1(\S)$ (based at $*$) onto the quotient
of the holonomy groups. These facts are proved in The Book, ref.
\cite{thebook}. We shall call the homomorphism $f$ the {\it homotopy map},
although a more accurate name would perhaps be ``homotopic holonomy map".

It follows from the results of the previous section that the association $f$
of a homotopy class of loops with an element of $\Phi/\Phi_0$ is
time-independent when the spin connection evolves according to the vacuum
Einstein equation. $f$ is also invariant under gauge transformations that are
the identity at $*$, and under diffeomorphisms that fix $*$ and are isotopic
to the identity. ``Large" diffeomorphisms, {\it i.e.,} ones that are not
isotopic to the identity,  can act on $\pi_1(\S)$ by a non-trivial
automorphism, so $f$ is not, in general, invariant under these.

According to the principle that one cannot physically distinguish
spacetimes that are related by a diffeomorphism, a true gravitational
observable must be invariant under large diffeomorphisms.\footnote{This
point of view has received support from an exact calculation of point-particle
scattering in $2+1$-dimensional quantum gravity \cite{scattering}.
In this setting, the large diffeomorphisms comprise a braid group, and Carlip
showed in \cite{scattering} that the correct semi-classical limit is obtained
only if one demands that the quantum state be invariant under this braid
group.}
Thus, the homotopy map is not quite an observable as it stands, even once $*$
and the gauge at $*$ are fixed. What is required is to factor out by the
action of the large diffeomorphisms. This leads one to the problem of
classifying such actions, which is not always an easy problem. Even if one
knows all the automorphisms of $\pi_1(\S)$, they are generally not all induced
by some diffeomorphism. Thus, there is no general solution to this problem.

To illustrate what is going on, let us consider a particular
case where the solution is known. For the manifold $\S=S^1\X R^2$, we have
$\pi_1(\S)=Z$, the additive group of integers. This group admits only
two automorphisms: the identity, and the mapping that sends $n$ to
$-n$. The inversion is induced by an inversion diffeomorphism that sends
$\theta$ to $-\theta$, where $\theta$ coordinatizes $S^1$. Thus, in this case,
we cannot associate an observable with an individual homotopy class, but
rather with a {\it pair} of such classes. The observable in question is
the corresponding {\it pair} of elements of $\Phi/\Phi_0$.

\vskip 1cm
\noindent {\bf 5. Classification}
\vskip .5cm

In this section the classification of possible homotopy maps
$f\!:\!\pi_1(\S)\rightarrow\Phi/\Phi_0$ will be obtained. This can be
accomplished in the following three stages:
\begin{enumerate}
\item classify $\Phi_0$;
\item classify $\Phi$ in which $\Phi_0$ is a normal subgroup, with
$\Phi/\Phi_0$ countable;
\item determine which of these $\Phi$'s can arise as holonomy groups
for a given 3-manifold $\Sigma$.
\end{enumerate}
In the next section, we shall add the requirement that the connection
occurs in a solution to the Einstein equation.

\vskip .5cm
\noindent{\sl Classification of $\Phi_0$}
\vskip .25cm

The only condition on the restricted holonomy group $\Phi_0$ is that it be a
connected Lie subgroup of $SL(2,C)$. In fact, we are interested in {\it
complex} subgroups, since it is these that are preserved in time.
These are determined by complex Lie subalgebras of $sl(2,C)$. As noted in
\cite{GLS}, all such subalgebras are equivalent via conjugation to one of
the following:
$$ sl(2,C), \qquad {\cal A}(+,3),\qquad {\cal A}(+),\qquad {\cal A}(3),
\qquad 0.$$
Here ${\cal A}(+,3)$, $\cA(+)$, and $\cA(3)$ denote the subalgebras generated
by $\{\t_+,\t_3\}$, $\t_+$, and $\t_3$, where
$$\tau_+=\pmatrix{0&1\cr0&0\cr}\quad{\rm and}\quad
\tau_3=\pmatrix{1&0\cr0&-1\cr}.$$
The corresponding groups $\Phi_0$ are $SL(2,C)$, the upper triangular
subgroup $G(+,3)$, the upper triangular subgroup with unit diagonal $G(+)$,
the diagonal subgroup $G(3)$, and the identity subgroup. The last case
corresponds to locally flat connnections.

\vskip .5cm
\noindent{\sl Classification of $\Phi$}
\vskip .25cm

As stated in section 4, $\Phi_0$ must be normal in $\Phi$. This is
because, in the loop group, the conjugate of a trivial loop by any
other loop is again trivial. Thus, for each case above, we shall find the
largest subgroup of $SL(2,C)$ in which $\Phi_0$ is normal
({\it i.e.,} the {\it normalizer} of $\Phi_0$), and then examine its subgroups.
Only those subgroups $\Phi$ for which the quotient $\Phi/\Phi_0$ is countable
can arise as possible holonomy groups, since the homotopy map $f$ is onto, and
the fundamental group $\pi_1(\S)$ is countable.
\vskip .2cm

\vskip .2cm
\noindent $\Phi_0=SL(2,C)$: We have $\Phi=\Phi_0$ and $\Phi/\Phi_0=\{id\}$.
\vskip .2cm

\vskip .2cm
\noindent $\Phi_0=G(+,3)$: One easily sees that $G(+,3)$ is its own normalizer.
Thus, again $\Phi=\Phi_0$ and $\Phi/\Phi_0=\{id\}$.
\vskip .2cm

\vskip .2cm
\noindent $\Phi_0=G(+)$: The normalizer of $G(+)$ is $G(+,3)$. The subgroups of
$G(+,3)$ that contain $G(+)$ are the groups of upper triangular matrices
with unit determinant whose diagonal components form a subgroup $K$ of the
non-zero complex numbers $C^*$. We will call these groups $G(+,K)$. For
$\Phi=G(+,K)$, one finds $\Phi/\Phi_0=K$. Thus, the possible $\Phi$'s in this
case are in one to one correspondence with countable subgroups of $C^*$. All
such subgroups are products of finite or infinite cyclic groups.
\vskip .2cm

\vskip .2cm
\noindent
$\Phi_0=G(3)$: The normalizer of $G(3)$ is the group consisting of all
diagonal and purely off-diagonal matrices (with unit determinant).
We shall call this group $G(3,Z_2)$. The only proper subgroup of $G(3,Z_2)$
that contains $G(3)$ is $G(3)$ itself, so $\Phi$ must be either all of
$G(3,Z_2)$ or $G(3)$. In the former case we have $\Phi/\Phi_0=Z_2$, where
$Z_2$ is the group of order 2. In the latter case, we have $\Phi/\Phi_0=
\{id\}$.
\vskip .2cm

\vskip .2cm
\noindent $\Phi_0=\{id\}$: The normalizer is all of $SL(2,C)$.  Thus,
$\Phi$ can be any countable subgroup of $SL(2,C)$, and $\Phi/\Phi_0=\Phi$.

\vskip .5cm
\noindent{\sl $\Phi$ as a holonomy group}
\vskip .25cm

The previous subsection provides a group-theoretic classification of those
$\Phi$ that may arise as the holonomy group of an $SL(2,C)$ connection on a
3-manifold $\S$.  If $\S$ is fixed, not all of these cases
will generally be realized by a connection.   A {\it necessary}
condition that $\Phi$ arise as a holonomy group is that there exist a
homomorphism (4) from $\pi_1(\S)$ onto $\Phi/\Phi_0$.  In fact, this is also
a {\it sufficient} condition, as will now be argued.

Let $f\!:\!\pi_1(\S)\rightarrow\Phi/\Phi_0$ be such a homomorphism, which we
assume exists. Choose a discrete subgroup $D\subset\Phi$ consisting of
one element from each coset of $\Phi/\Phi_0$, so that $\Phi/\Phi_0\rightarrow
D$
is an isomorphism of groups.\footnote{This is possible for all $\Phi$
associated with $SL(2,C)$.  In general,  it is possible if and
only if $\Phi$ is isomorphic to the semidirect product of $\Phi/\Phi_0$ with
$\Phi_0$.} Then we can define a
locally flat connection with holonomy group $D$ by lifting $\pi_1(\S){\buildrel
f\over\rightarrow}\Phi/\Phi_0\rightarrow D$ to a
holonomy map from the loop group to $D$.  This is because any homomorphism
from the loop group to the structure group, that vanishes on so-called ``thin"
loops, determines a principal bundle and connection up to gauge transformations
\cite{barrett}.\footnote{A smoothness condition must also be assumed in order
that the bundle and connection be differentiable. This condition is
satisfied for the holonomy map defined above.} To obtain
a connection with holonomy group $\Phi$, one need only make a generic
deformation of the connection 1-form taking values in the Lie algebra of
$\Phi_0$. The curvature will then ``fatten" the holonomy of a homotopy class
associated with $g\in D$ to the corresponding coset $g\Phi_0$.  To be fancy,
one can invoke the Ambrose-Singer theorem \cite{thebook}, which states that
the holonomy algebra at a point $p$ in the principal fiber bundle
is generated by the spans of the curvature 2-forms on the holonomy bundle
through $p$. Thus, any deformation of the connection for which the local
curvature spans the Lie algebra of $\Phi_0$ will yield the holonomy group
$\Phi$.

Do manifolds $\S$ and (onto) homomorphisms $f\!:\!\pi_1(\S)\rightarrow
\Phi/\Phi_0$ always exist?  The answer is yes. For suppose that the fundamental
group of $\S$ is a free group on $n$ generators, with $n$ greater
than or equal to the number of generators of $\Phi/\Phi_0$.
(For instance, $\Sigma$ could be a 3-sphere with $n$ handles.)
Then since
there are no relations among the generators of $\pi_1(\S)$, one can define
$f$ by freely assigning at least one generator of $\pi_1(\S)$ to each
generator of $\Phi/\Phi_0$, and then extending this map to a homomorphism.

Thus, each $\Phi$ classified in the previous subsection can arise as the
holonomy group of an $SL(2,C)$ connection on {\it some} 3-manifold $\S$.
Indeed,
all the $\Phi$'s can occur on the same manifold $\S$, provided the fundamental
group of $\S$ is free and has an infinite number of generators (although
this probably entails $\S$ being non-compact).

\vskip .5cm
\noindent{\sl Examples}
\vskip .25cm

In this subsection, some examples of $SL(2,C)$ connections that have
disconnected holonomy groups will be given. These serve to make the preceeding
classification more concrete. As noted above, the only disconnected holonomy
groups for $SL(2,C)$ connections that are not locally flat are $(i)\; \;
\Phi=G(+,K)$ where $K$ is any countable subgroup of $C^*$, and $(ii)\; \;
\Phi=G(3,Z_2)$.  We now take these in turn.

\vskip .5cm
\noindent ($i$) $\Phi=G(+,K)$
\vskip .25cm

If in a global gauge the connection $\w$ is of the form
$\w=\w^+\t_+ + \w^3\t_3$, then the holonomy $h_\g[\w]$ around a loop $\gamma$
is an element of $G(+,3)$ given by
$$h_\g[\w]=\pmatrix{\exp\oint_\g\w^3&z_\g\cr0&\exp(-\oint_\g\w^3)\cr}.
\eqno(5a)$$
Here
$$z_\g=(\exp\oint_\g\w^3)\;\oint_\g \ e^{-2\int_0^s\w^3}\;\w^+,
\eqno(5b)$$
and $s$ denotes a parameter along the loop $\g$. In order to restrict
the connection so that $\Phi_0\subset G(+)$, it is sufficient
that the 1-form $\w^3$ be closed, so that $\oint_\g\w^3=0$ when $\g$ is
homotopic to the identity.  For a generic $\w^+$, $\Phi_0$ will then comprise
all of $G(+)$.

Now let $K$ be a countable subgroup of $C^*$, and let $\{a_i\}$ denote a
(possibly infinite) set of non-zero complex numbers that generate $K$.
Suppose that the number of generating elements of $K$ is less than or equal to
the dimension of the first homology group of $\S$, $H_1(\S)$, and let $\{\g_i
\}$ denote a generating set of 1-cycles for $H_1(\S)$.\footnote{We need to
introduce homology groups in order to later appeal to de Rahm's theorem.} To
arrange for the holonomy group to be $G(+,K)$, it suffices to impose the
further restrictions that the  closed 1-form $\w^3$ satisfy $\oint_{\g_i}\w^3
=\ln a_i$.  If there are more generators $\g_i$ than $a_i$, then we can
require that $\oint_{\g_i}\w^3=0$ for the additional $\g_i$.
The existence of such a closed 1-form is guaranteed by the de Rahm
theorem (see, {\it e.g.,} \cite{threeladies}), which establishes the
duality of the first homology and cohomology groups.  Note
that by defining $\w$ in this fashion, we have effectively encoded all the
``extra holonomy'' in the $\t_3$ part of the connection.

For example, suppose that the manifold $\S=T^2\X R$,
and $K$ is generated by two generic non-zero complex numbers $a_1$ and $a_2$.
$K$ is then isomorphic to $Z\X Z$, unless $a_1$ or $a_2$ is a root of unity.
The manifold $T^2\X R$ has both its fundamental and first homology groups equal
to $Z\X Z$. Let us take $\g_1=(\theta_1,0,0)$ and $\g_2=(0,\theta_2,0)$ as the
two generating loops on $\S$, where $\theta_1,\theta_2\in[0,2\pi)$ coordinatize
the torus. Then the above discussion shows that the connection
$$\w=\w^+ \;\t_+ + {1\over 2\pi}(\ln a_1 \; d\theta_1+\ln a_2 \;
d\theta_2)\;\t_3,\eqno(6)$$
has holonomy $\Phi=G(+,Z\X Z)$, for $\w^+$ any non-vanishing 1-form on $R$.
In particular, one could choose $\w^+=dz$, where $z$ is a coordinate
on $R$ that is constant on $T^2$. In this case, the curvature that ``fills out"
$\Phi_0=G(+)$ comes from the commutator $[\t_3,\t_+]=2\t_+$.

The connections above have been defined in a global gauge, and take values in
the algebra ${\cal A}(+,3)$, although the holonomy algebra itself is
${\cal A}(+)$. This means that in a covering family of radial gauges
(see section 4 and the Appendix), the connection takes values only in
${\cal A}(+)$, and the gauge transition functions take values in $G(+,3)$.
This observation may be important if one tries to construct solutions with
non-trivial homotopy observables.

\vskip .5cm
\noindent ($ii$) $\Phi=G(3,Z_2)$
\vskip .25cm

Instead of working in a global gauge like we did in
the previous example, we will take $\S=S^1\X S^2$, and cover $\S$ with two
local gauges $\s_0$ and $\s_1$ defined on $U_0\X S^2$ and $U_1\X S^2$. (Here
$U_0$ and $U_1$ cover $S^1$, with two overlap regions.)
These gauges are chosen so that they agree on one overlap region, but disagree
on the other, where they are related by the constant transition function
$\psi_{01}={\scriptsize{\pmatrix{0&1\cr-1&0\cr}}}$.

Suppose that in each local gauge the connection is of the form $\w=\w^3\;\t_3$.
Then since the transition function $\psi_{01}$ lies in the normalizer
of $G(3)$, we have $\Phi_0\subset G(3)$.\footnote{This is a special case
of a general result:  If a connection takes values in a subalgebra $h$ in each
gauge patch, and if the transition functions lie in the normalizer of the
corresponding subgroup $H$, then the holonomy around any contractible
loop lies in $H$.}  For generic $\w^3$, $\Phi_0$ will comprise all of $G(3)$.
To see that $\Phi=G(3,Z_2)$, let $\g$ be a loop that wraps once around $S^1$
and traverses each overlap region once. Since $\psi_{01}$ lies in the
normalizer of $G(3)$, equation (3) shows that the holonomy $h_\g[\w]$ is a
product of $\psi_{01}$ with some element of $G(3)$.  This product belongs to
the non-identity coset in $G(3,Z_2)$. Since $\Phi_0$ is all of $G(3)$, $\Phi$
is thus all of $G(3,Z_2)$.

Since the local gauge $\s_1$ can be deformed to agree with
$\s_0$ in both overlap regions, one can write the connection (defined above)
in a global gauge without the need of transition functions.  But in order to
do this, one would have to perform a gauge transformation that is the
identity in one overlap region, and takes $\psi_{01}$ to the identity in the
other. This gauge transformation would necessarily involve group elements of
the form $\exp(\a\psi_{01})$, where $\a\in[0,2\pi)$.  That the holonomy group
equals $G(3,Z_2)$ would no longer be manifest if the connection were expressed
in this global gauge.

\vskip 1cm
\noindent {\bf 6. The Einstein equation}
\vskip .25cm

A classical observable is a function on the space of solutions to the
equations of motions modulo gauge transformations. Thus, only those
holonomy groups that arise in a Lorentzian solution to the Einstein
equation are relevant. It is thus of interest to determine which holonomy
groups classified above actually occur in solutions, and to characterize
those solutions as far as possible.\footnote{The relevance of
this issue goes beyond the holonomy group considerations of the present
paper.  It would seem to be of fundamental interest to the Ashtekar variables
program to have the general answer, in order to know which connections
actually lie in the reduced phase space. In the quantum theory, those
connections that are not in the reduced phase space should be associated with
zero ``measure" in a functional integral or Hilbert space inner product.}

In the first subsection, {\sl Local solutions}, we will consider only the
{\it local} restrictions imposed by the Einstein equation, for each of the
possible holonomy subalgebras. These local considerations do not suffice,
however, to establish (or classify) the existence of globally regular solutions
with a given holonomy group. One would ideally like to classify such solutions
for each 3-manifold $\S$ and each holonomy group $\Phi$. In the second
subsection, {\sl Global considerations}, this problem will be discussed.
We have no complete answer to this global problem as of now. However, some
partial results will be given.

\vskip .5cm
\noindent{\sl Local solutions}
\vskip .25cm

If the holonomy algebra of a spacetime is not
all of $sl(2,C)$, then it is contained in $\cA(+,3)$, and there is
a covariantly constant null {\it direction} in the spacetime.
The vacuum solutions with this property are the Goldberg-Kerr
solutions \cite{GK}. These have a line element that can be locally written in
the form [1]
$$ds^2=2dz d\ovr z-2du(dr+Wdz+\ovr Wd\ovr z+H du),\eqno(7a)$$
where
$$\eqalignnotwo{&H={1\over 2}(W_{\ovr z}+\ovr W{}_z)r + H^0,&(7b)\cr
&H^0=Re[(W W_{\ovr z}+W_u)z + h(z,u)],&(7c)\cr}$$
and $W(\ovr z,u)$, $h(z,u)$ are arbitrary functions holomorphic in $\ovr z$
and $z$, respectively. (Subscripts denote partial differentiation.)

If the holonomy algebra is $\cA(+)$, then there is a covariantly constant
null {\it vector}, and the only vacuum solutions with this property are the
{\sl pp} waves \cite{exact}. These correspond to the Goldberg-Kerr solutions
above with $W=0$.
If the holonomy algebra is ${\cal A}(3)$, then as shown
in \cite{GLS} (using Ashtekar's form of the initial value
constraints), the only vacuum solution is flat space, so the algebra is
actually trivial. Thus, the only vacuum solutions with non-trivial
homotopy map are spacetimes that are locally {\sl pp} waves with holonomy
group $G(+,K)$, and locally flat spacetimes.

In the presence of a cosmological constant, the locally flat solutions
are excluded. In addition, Bostr\"om has shown \cite{bostrom}
(again, using the Ashtekar constraints)
that there are no solutions with holonomy algebra $\cA(+)$.
To our knowledge, the solutions with non-zero cosmological constant and
holonomy algebra equal to $\cA(+,3)$ have not been classified.

The remaining case to consider is that of the holonomy algebra $\cA(3)$.
It turns out that this case is allowed in the
presence of a cosmological constant $\L$, although there is locally only one
solution for each value of $\L$. To see this, note first that there are now
{\it two} covariantly constant null directions, say those of the vectors
$l^\u$ and $n^\u$. It is easy to verify that $l^\u$ and $n^\u$ are both
hypersurface orthogonal and surface-forming. We can therefore introduce
null coordinates $u$, $v$ and complex spatial coordinates $z$, $\ovr z$
for which the line element takes the form
$$ds^2=|f|^2(u,v,z,\ovr z)\;  dz d\ovr z-g(u,v,z,\ovr z)\; du dv.\eqno(8)$$
The condition that the holonomy algebra is $\cA(3)$ implies that the
connection generates only boosts in the $u,v$-plane and rotations in the
complex $z$-plane, which in turn implies $|f|^2=|f|^2(z,\ovr z)$ and
$g=g(u,v)$.
The spacetime is thus locally a direct product of 2-dimensional Euclidean
and Lorentzian spaces.

Applying the Einstein equation with cosmological constant ($R_{\u\v}=\L
g_{\u\v}$), we find that the 2-dimenstional sections are spaces of the same
constant curvature. If $\L=0$ we have flat space; if $\L > 0$ we have the
product of a 2-sphere with 2-dimensional de Sitter space; and if $\L < 0$ we
have the product of a 2-hyperboloid with 2-dimensional anti-de Sitter space.
These solutions with $\L\not=0$ are the Nariai solutions \cite{nariai}.

Thus the {\it only} solutions with $\Lambda\ne0$ and non-trivial homotopy map
are locally the Nariai solutions, with holonomy group $G(3,Z_2)$.

\vskip .5cm
\noindent{\sl Global considerations}
\vskip .25cm

There are two types of global issues to consider: classification of
globally regular solutions with a given holonomy {\it algebra}, and
classification of globally regular solutions with a given disconnected
holonomy {\it group}.

Let us consider first the latter question, in the case where the connection is
locally flat. For locally flat connections, the problem is closely analogous
to one that has been answered recently for flat $SO_0(2,1)\cong PSL(2,R)$
connections on 2-manifolds, $\Sigma^2$ \cite{mess,measuring}.
In that case, if the genus of $\Sigma^2$ is greater than one, a
compatible triad for which $\Sigma^2$  is spacelike exists if and only if
the homotopy map is a discrete embedding of $\pi_1(\Sigma^2)$ into
$SO(2,1)$.  The $3+1$-dimensional case is somewhat different, not only
because of the added dimension, but also because one is specifying just the
self-dual part of the 4-dimensional spin-connection. Nevertheless,
according to Carlip \cite{carlippc}, a similar result holds. Namely, if $G$ is
a discrete subgroup of $SL(2,C)$, then $G$ acts properly discontinuously on the
hyperboloids of constant proper time that foliate the interior of a light cone
$X$ in $3+1$ dimensional Minkowski spacetime. The quotient of $X$ by $G$ is a
flat spacetime with induced spin connection whose holonomy group is $G$.
The question of whether a flat spacetime with a discrete holonomy group must
arise in this manner remains open, as does the question of whether
the holonomy group must be discrete.

The Nariai solutions mentioned above for $\L\not=0$ are the only solutions
with holonomy algebra $\cA(3)$.  There are various globally regular forms
of these metrics, obtained by identifications.  For example, for $\L < 0$,
quotients of the
hyperbolic 2-space will yield the Teichm\"uller spaces of metrics on genus
$g\geq 2$ surfaces.  The holonomy group of all of these spacetimes will be
$G(3)$.

Is it is possible to patch together local Nariai metrics in such a way as to
obtain globally regular solutions with the disconnected holonomy group
$G(3,Z_2)$?  The answer seems to be no. For instance, consider the $\L > 0$
Nariai solution and the $\Phi=G(3,Z_2)$ example of section 5.
If $\S$ of that example is identified with
a spatial slice of the Nariai solution, then the $SL(2,C)$ transition
function $\psi_{01}={\scriptsize{\pmatrix{0&1\cr-1&0\cr}}}$ corresponds to
a rotation through $90^\circ$ in a plane orthogonal to the $S^2$-tangent
plane.  If we were to try patching together local solutions using such a
transition function, we would ``mix up" the direct product structure of the
Nariai metric.  The transition function thus could not be induced by an
isometric coordinate tranformation, so such a patching seems to be impossible.

Finally, consider the solutions that have holonomy algebra $\cA(+,3)$,
that is, the Goldberg-Kerr solutions (7).
The 2-dimensional spaces $r,u={\rm const}$ are {\it flat}, so
if orientable, they must be the plane, the cylinder, or the torus $T^2$.
Global regularity imposes severe restrictions on the $z$-dependence of the
holomorphic functions $W$ and $h$ in the line element.  For the $T^2$
case, the only holomorphic function is a {\it constant}, so $W(\ovr z,u)=
W(u)$ and $h(z,u)=h(u)$.  In fact, $W$ must also be independent of $u$,
since otherwise the linear term in $z$ in $H^0$ will not be globally
extendible.  The flat metric on the torus can have any Teichm\"uller
parameters.

As discussed in the previous subsection, the {\sl pp} wave solutions ($W=0$)
have holonomy algebra $\cA(+)$.  Are there any
gobally defined solutions that are locally {\sl pp} waves, and that have a
disconnected holonomy group of the form $G(+,K)$?  That is, can they support
global holonomy in a discrete subgroup of $G(3)$?  We do not know the answer
to this question.  However, we can offer the following remark.

Recall that the homotopy map $\pi_1(\S)\rightarrow\Phi/\Phi_0=K$ is {\it onto},
so a non-trivial $K$ requires a non-trivial $\pi_1(\S)$.  For instance, if $\S=
T^2\X R$, then $K$ can have at most two generators.  One might think that
non-trivial Teichm\"uller parameters on $T^2$ would in general lead to global
holonomy of the form $G(+,Z\X Z)$, but this is not the case.  In
2+1-dimensions, Carlip \cite{observables} has shown explicitly how the
Teichm\"uller parameters are expressed in terms of the generators of two
commuting $ISO(2,1)$ holonomies. The homogeneous parts of these holonomies are
boosts in directions tangent to the torus.  Embedding Carlip's analysis in
3+1-dimensions, we find that
these boosts cannot be identified with boosts in $G(3)$, because the latter
boost in the spatial direction orthogonal to the torus.\footnote{The null
rotations in $G(+)$ are generated by $\t_+=(\t_1+i\t_2)/2$. In an adapted
spin-frame, these spin-transformations stabilize the covariantly constant
null vector, which is orthogonal to the torus. Therefore, in the adapted
spin-frame, it is the 1- and 2-axes that are tangent to the torus.  The 3-axis
is thus orthogonal to the torus.} Thus, to obtain {\sl pp} wave solutions with
global holonomy in $G(3)$, it is evidently necessary to patch together local
solutions with transition functions in $G(3)$.  We do not know whether or not
this is possible.

\vskip 1cm
\noindent{\bf 7. Discussion}
\vskip .5cm

The conservation of the holonomy group (or, more generally, the homotopy map)
of the chiral spin-connection in vacuum GR may turn out
to have interesting applications. At the classical level, it provides a global
constraint on evolution of initial data, that can only be violated due to
the occurence of a Cauchy horizon or singularity. Put differently, data sets
with different holonomy groups cannot be cobordant in a regular vacuum
solution.

For use in the quantum theory, the major drawback is probably that
the holonomy group is all of $SL(2,C)$ for a generic solution, so
the holonomy group observable does not distinguish typical solutions.
However, in a ``midi-superspace" consisting of connections with holonomy
less than all of $SL(2,C)$, it would be relatively more significant.
It might  be interesting to formulate a midi-superspace quantization of general
relativity along these lines, as an alternative to the usual truncations of
the theory.

That the holonomy group is not conserved in the presence of matter couplings
is, of course, another limitation in its applicability. One way to evade this
limitation would be to model matter by vacuum configurations such as the
Einstein-Rosen bridge of the extended Schwarzschild solution. However, since
the Schwarzschild solution has holonomy group equal to all of $SL(2,C)$, this
is not so useful. Another option would be to couple gravity to one-dimensional
string-like matter. Then the energy-momentum tensor would vanish everywhere
except on the world sheets,
where it would be singular. If the worldsheets are removed from the spacetime,
what remains is a vacuum solution, so the holonomy group should be conserved.
For special string configurations, the holonomy group can be less than all of
$SL(2,C)$. If the strings are knotted or linked, interesting homotopy groups
can arise, allowing for a rich array of holonomy groups and their associated
homotopy maps.

\vskip 1cm
\noindent{\bf Appendix: Reduction Theorem}
\vskip .5cm

The reduction theorem played such a crucial role in establishing the
time-independence of the holonomy group that it seems worthwhile to illustrate
the logic underlying this theorem. Since the principal bundle point of view
is not familiar to all physicists, we shall just give a ``low brow'' proof
that $\S$ can be covered by a collection of local gauge patches such that
($a$) in each gauge patch the connection takes values in the holonomy algebra
based at $*$,
and ($b$) the gauge tranformations relating the patches take values in the
holonomy group based at $*$. This result is valid for {\it any} $G$-bundle
over {\it any} connected $n$-manifold $\S$.

The argument is simplest when $\S$ is contractible, because then we can
construct a single, globally defined {\it radial gauge} based at $*$, in
which the connection takes values in the holonomy algebra based at $*$.
This is done as follows: Choose an arbitrary family of curves through $*$
that never cross and that fill all of space. Then choose an arbitrary gauge
at $*$, and carry it out to the rest of $\S$ by parallel transport along
these curves.  This defines a global gauge, in which the component
of the connection along these curves vanishes.  This means that if a
loop is formed by segments of two of these curves joined by a transverse piece,
the only non-trivial contribution to the holonomy will come from the transverse
piece. If the transverse piece is infinitesimal, this contribution to the
holonomy differs from the identity by a term proportional to the connection.
Thus, in a radial gauge based at $*$, the connection takes values in
the holonomy algebra based at $*$.

If $\S$ is not contractible, then a global radial gauge does not exist.
However, we can cover $\S$ by a collection of local radial gauge patches
satisfying
conditions ($a$) and ($b$) as follows:  Choose a fiducial gauge at $*$, and
carry it out to a collection of points $\{x_i\}$ (whose local gauge patches
will cover $\S$) by parallel transport along a collection of
curves $\{\g_i\}$. At each $x_i$, extend the resulting gauge to a local radial
gauge $\s_i$, with one of the radial curves being $\g_i{}^{-1}$ (the curve
$\g_i$ traversed in the opposite direction).
All of the local gauges $\s_i$ cover the point $*$ and agree with the
fiducial gauge there. Since the component of the connection along all the
$\g_i$ vanishes in the gauge $\s_i$,
the holonomy group based at $x_i$ in the gauge $\s_i$ coincides with
the holonomy group based at $*$ in the fiducial gauge.
In fact, the holonomy group based at
{\it any} point $y$ in any local gauge $\s_i$ agrees with the holonomy
group based at $*$ in the fiducial gauge. This is because
$y$ and $x_i$ are always joined by some radial curve $\g_{iy}$ of the gauge
$\s_i$, so the component of the connection along the combined
curve $\g_i\g_{iy}$ (which connects $*$ to $y$) vanishes.

To show that condition ($a$) holds, we adopt the collection of local radial
gauges
$\{\s_i\}$ defined above, and apply the argument used previously in the
contractible case, but now with the holonomy based at $x_i$.
In this set of gauges, the connection indeed takes values in the holonomy
algebra based at $*$. To show that condition ($b$) holds,
we consider any point $y$ in the overlap of two gauges $\s_i$ and $\s_j$.
The gauge transformation at $y$ from the gauge $\s_i$ to the gauge $\s_j$
is given by the holonomy at $y$ around the loop $\g_y:=\g_{yi}\g_i{}^{-1}\g_j
\g_{jy}$, evaluated in the gauge $\s_i$. But as explained above,
the holonomy group based at $y$ in the gauge $\s_i$ coincides with the
holonomy group based at $*$ in the fiducial gauge. Thus, the gauge
transformation is indeed an element of the holonomy group based at $*$.
\vskip 1cm

\noindent ACKOWLEDGEMENTS

\vspace{.3cm}

We are grateful to Robert Bryant, Steve Carlip, Bill Goldman, Jerzy
Lewandowski, Charlie Misner, and Chris Stark for helpful conversations.
This research was supported in part
by NSF grant PHY91-12240.

\end{document}